\documentclass[copyright,creativecommons]{eptcs}
\providecommand{\event}{WACA 2025} % Name of the event you are submitting to

\usepackage{iftex}

\ifpdf
  \usepackage{underscore}         % Only needed if you use pdflatex.
  \usepackage[T1]{fontenc}        % Recommended with pdflatex
\else
  \usepackage{breakurl}           % Not needed if you use pdflatex only.
\fi
\usepackage{tabularx}
\usepackage{graphicx}
\usepackage{hyperref}
\usepackage{amsmath}
\usepackage{amssymb}
\usepackage{booktabs}
\usepackage{url}
\usepackage[english]{babel}
\PassOptionsToPackage{strings}{underscore}
\usepackage[strings]{underscore}
\usepackage{listings}
\usepackage{xcolor}
\usepackage{paralist}
\usepackage{pifont} % for ding symbols

\newcommand{\cmark}{\checkmark}
\newcommand{\xmark}{\ding{55}}
\newcommand{\pmrk}{\(\sim\)} % "partial" marker

\definecolor{javared}{rgb}{0.6,0,0} % for strings
\definecolor{javagreen}{rgb}{0.25,0.5,0.35} % for comments
\definecolor{javapurple}{rgb}{0.5,0,0.35} % for keywords
\definecolor{javadocblue}{rgb}{0.25,0.35,0.75} % for javadoc comments

\providecommand{\thisvolume}[1]{this volume of EPTCS, Open Publishing Association}

\lstdefinestyle{java}{
  language=Java,
  basicstyle=\ttfamily\small,
  keywordstyle=\color{javapurple}\bfseries,
  stringstyle=\color{javared},
  commentstyle=\color{javagreen}\itshape,
  morecomment=[s][\color{javadocblue}]{/**}{*/},
  numbers=left,
  numberstyle=\tiny\color{gray},
  stepnumber=1,
  numbersep=10pt,
  backgroundcolor=\color{white},
  showspaces=false,
  showstringspaces=false,
  showtabs=false,
  frame=single,
  tabsize=2,
  captionpos=b,
  breaklines=true,
  breakatwhitespace=false,
  escapeinside={(*@}{@*)}
}

\lstdefinestyle{yaml}{
  language=,
  basicstyle=\ttfamily\small,
  keywordstyle=\color{blue},
  commentstyle=\color{gray},
  breaklines=true,
  frame=single,
	numbers=left,
	numberstyle=\tiny\color{gray},
  stepnumber=1,
  numbersep=10pt,
  backgroundcolor=\color{white},
  showspaces=false,
  showstringspaces=false,
  showtabs=false
}

\lstdefinestyle{gocode}{
  language=Go,
  basicstyle=\ttfamily\small,
  keywordstyle=\color{blue},
  commentstyle=\color{gray},
  breaklines=true,
	numbers=left,
  frame=single,
	numberstyle=\tiny\color{gray},
  stepnumber=1,
  numbersep=10pt,
  backgroundcolor=\color{white},
  showspaces=false,
  showstringspaces=false,
  showtabs=false
}

\title{Decoupling Adaptive Control in TeaStore}

\author{Eddy Truyen
\institute{DistriNet, KU Leuven, 3001 Leuven, Belgium}
\email{Eddy.Truyen@kuleuven.be}}

\def\titlerunning{Decoupling Adaptive Control in TeaStore}
\def\authorrunning{E. Truyen}

\begin{document}

\maketitle

%-------------------------------------------------------------------------------
\begin{abstract}
The Adaptable TeaStore specification provides a microservice-based case study for implementing self-adaptation through a control loop. 
We argue that implementations of this specification should be informed by key properties of self-adaptation: \emph{system-wide consistency} (coordinated adaptations across replicas), \emph{planning} (executing an adaptation until appropriate conditions are met),  and \emph{modularity} (clean integration of adaptation logic). 
In this implementation discussion paper, we examine how software architectural methods, the cloud-native Operator pattern, and legacy programming language techniques can decouple self-adaptive control logic from the TeaStore application. We analyze the trade-offs that these different approaches make between fine-grained expressive adaptation and system-wide control, and highlight when reuse of adaptation strategies is most effective. Our analysis suggests that these approaches are not mutually exclusive but can be combined into a multi-tiered architecture for self-adaptive microservices.
\end{abstract}

\section{Introduction}

Self-adaptive software systems have been a long-standing research focus for several decades~\cite{1160055, 1301377, weyns2020introduction}. A key conceptual foundation is the MAPE-K control loop~\cite{1160055}, which stands for \emph{Monitoring, Analysis, Planning, Execution with Knowledge}. This model has guided research in areas ranging from software engineering~\cite{1301377, weyns2020introduction} to distributed systems~\cite{porter2016rex}.

With the rise of cloud-native computing, self-adaptation has become increasingly critical. Autoscaling of microservices for performance efficiency is a well-known example. In contrast, \emph{self-adaptive behavior within microservices} has received less attention. To address this gap, the \emph{Adaptable TeaStore} specification~\cite{BDGLZZ25} proposes a benchmark for fine-grained adaptation within components. It envisions configuring MAPE-K control loops inside individual microservices by defining rules that detect event patterns, decide whether adaptation is needed, and trigger effectors to adjust internal behavior. Since the specification itself is abstract, implementation-oriented studies are required to explore how fundamental self-adaptation properties -- such as planning, system-wide consistency, and modularity -- can be realized in practice.

\subsection{Adaptable TeaStore specification}

The Adaptable TeaStore specification proposes to define a self-adaptive control loop to enable fine-grained and tailored adaptation of microservices. The specification adapts the MAKE-K control loop but monitoring and effector interfaces are directly implemented within the code of the microservices. Thus, it provides the M and E phases of the MAPE-K loop. All decisions related to the A and P phases (and to the knowledge management) are left open. Effectors may directly change the behavior of a microservice. The desired configuration to be applied by an effector can also be exposed through a REST API~\cite{BDGLZZ25}. In addition, the specification proposed a number of self-adaptation scenarios that can be implemented using different approaches. One of these scenarios is switching the TeaStore application to low-power mode in circumstances of user request overload or a Denial-of-Service (DoS) attack. For the Recommender microservice, switching to low-power mode entails deploying a more light-weight Machine Learning (ML) model.    

\subsection{Motivation}

The Adaptable TeaStore specification aims to embed monitoring and effector interfaces within the individual microservices. However, this approach raises questions when compared to well-understood properties of self-adaptation. In particular, achieving (1) \emph{system-wide consistency} (coordinated adaptations of microservice replicas), (2) \emph{modularity} (reusable and non-invasive adaptation logic), and (3) \emph{planning} (delaying adaptation until appropriate conditions are met) may require additional mechanisms for control and coordination:

\begin{enumerate}
    \item \textbf{System-wide consistency.} A naive implementation, which also implements the A and P phases within the code of the microservices (e.g., a rule-based system of which the rules are specifically designed for that microservice), leads to local changes to a single microservice instance (e.g., a container) that are difficult to apply consistently at the level of a \emph{microservice deployment} with multiple replicas. For example, if one replica switches to low-power mode when the average response latency exceeds a threshold, the same adaptation should be applied across all replicas to ensure consistent behavior as perceived by clients or other services.
  
	 \item \textbf{Planning.} In the planning phase of the MAPE-K loop, it is important to treat deployment and activation of an adaptation as separate steps. For example, when the Recommender microservice switches to a lightweight ML model, incoming requests may fail until that model has finished training. One way to avoid this is through a dark launch: the new model is deployed in the background but not yet exposed to clients while it trains. The downside is that the system cannot switch to low-power mode immediately. If an instant switch to low-power mode is required, the system must fall back to a default, non-ML-based recommender until the ML model becomes ready. This illustrates that planning is complex: several options may need to be deployed and prepared in advance, but only one will actually be activated at run time.
   
	\item \textbf{Modularity.} Behavioral variations within a microservice are often hard to modularize. Static variables that influence the selection of class instances in design patterns such as Strategy or Factory may lead to code duplication, since alternative classes can mix base functionality with variation-specific logic. Advanced modularization techniques may help improve separation of concerns and reduce duplication. For instance, additional exception handlers could be modularly integrated when adaptation actions interfere with internal microservice behavior. Likewise, the entire MAPE-K control loop should be designed for modular integration so it can be automatically applied to new releases of the original TeaStore benchmark.

\end{enumerate}

In the remainder of the paper, we will use system-wide consistency, planning, and modularity as key lenses for analyzing different techniques.

\subsection{Contribution}
This paper makes three main contributions:

\begin{enumerate}
    \item It introduces two approaches for using a REST API to achieve system-wide consistency and planning during self-adaptation: an architecture-based adaptation strategy and the Kubernetes Operator pattern.
    \item It reviews programming language techniques that improve the modular integration of self-adaptive properties into existing microservice applications, focusing on Aspect-Oriented Programming (AOP) and Context-Oriented Programming (COP).
    \item It analyses how these four approaches apply to the scenarios of the Adaptable TeaStore specification and identifies missing specification elements that are essential for robust, reusable, and comparable self-adaptive implementations. This provides concrete guidance for future extensions of the Adaptable TeaStore benchmark and for the design of multi-tiered adaptation frameworks.
\end{enumerate}

\subsection{Structure of the paper}
First, Section \ref{sect:background} introduces the four approaches and discusses their advantages and disadvantages with respect to the three aforementioned properties for self-adaptation. Then Section \ref{sect:operatorpattern} applies the Operator pattern in the context of the Adaptable TeaStore specification, focusing on system-wide consistency and planning. Thereafter, Section \ref{sect:modularity} shows how the programming languages for advanced modularity can be used to separate the control loop from the code of microservices and how to dynamically activate adaptations in the scope of specific client requests. Finally, Section \ref{sect:conclusion} discusses the main lessons learned.

\section{Background}
\label{sect:background}

Alternative approaches such as architecture-based adaptation, the Operator pattern and programming languages for advanced modularity provide different means to externalize or modularize self-adaptation. This section introduces these methods and techniques with a focus on their known advantages and disadvantages. Table~\ref{tab:comparison} summarizes the main advantages and disadvantages of each approach.

%\begin{table}[h]
%\centering
%\begin{tabularx}{\textwidth}{@{}l X X@{}}
%\toprule
%\textbf{Technique} & \textbf{Advantages} & \textbf{Limitations} \\ \midrule
%Architecture-based adaptation & Generic support for planning and system-wide consistency & Not reusable across applications (without a standard platform like K8s)\\  
%Operator pattern & System-wide consistency, planning to some extent & Development complexity, lack of modular reasoning \\ 
%AOP & Separation of concerns & Lack of modular reasoning, no system-wide consistency in distributed systems \\
%COP & Dynamic user-specific adaptation, modularity, consistency & Scope management complexity, challenging in asynchronous/event-based systems \\ 
%\bottomrule
%\end{tabularx}
%\caption{Comparison of approaches for decoupling self-adaptive control}
%\label{tab:comparison}
%\end{table}

\begin{table}[h]
\centering
\begin{tabularx}{\textwidth}{@{}l *{4}{>{\centering\arraybackslash}X}@{}}
\toprule
\textbf{Property} &
\textbf{Architecture-based} &
\textbf{Operator pattern} &
\textbf{AOP} &
\textbf{COP} \\ \midrule

System-wide consistency in distributed systems &
\protect\cmark & \protect\cmark & \protect\xmark & \protect\xmark \\

Support for planning &
\protect\cmark & \protect\pmrk & \protect\xmark & \protect\xmark \\

Reusability across applications &
\protect\xmark & \protect\cmark & \protect\cmark & \protect\cmark \\

Separation of concerns &
\protect\pmrk & \protect\pmrk & \protect\cmark & \protect\cmark \\

Modular reasoning &
\protect\pmrk & \protect\xmark & \protect\xmark & \protect\xmark \\

Dynamic adaptation at runtime &
\protect\cmark & \protect\cmark & \protect\pmrk & \protect\cmark \\

Complexity / scope management &
\protect\pmrk & \protect\xmark & \protect\pmrk & \protect\xmark \\

\bottomrule
\end{tabularx}

\vspace{0.5em}
{\small
\textbf{Legend:}  
\protect\cmark\ = fully supported \quad
\protect\xmark\ = not supported \quad
\protect\pmrk\ = partially supported / conditional
}

\caption{Comparison of approaches for decoupling self-adaptive control}
\label{tab:comparison}
\end{table}

\subsection{Architecture-based adaptation}

The Rainbow framework separates adaptation concerns into an external control architecture based on the MAPE-K model. This control layer abstracts the underlying system through two interfaces: one for monitoring (observation) and one for executing adaptations (effectors)~\cite{1301377}.

A retrospective study of Rainbow noted that the framework requires development teams to construct the entire system architecture from scratch, including the control logic, probes, effectors, and the application itself~\cite{generalityvsreusability}. While this design makes Rainbow broadly applicable, it also limits reusability---porting it to a new application often demands significant redevelopment.

Cloud-native computing, and Kubernetes in particular, offers a way to address this trade-off. Kubernetes (K8s) is the de-facto standard for container orchestration in cloud computing and its API is supported by all major cloud providers~\cite{Kubernetes, fi15020063}. By leveraging K8s as a common cloud-native platform, the Rainbow framework can achieve both generality and reusability~\cite{generalityvsreusability}. Indeed, self-adaptive frameworks such as k8-scalar and Kubow build on the REST API of K8s and commonly deployed monitoring subsystems. This allows them to define custom control loops that can be reused across any K8s-based application~\cite{delnat2018k8,kubow}.

Consider, for example, the Rainbow framework applied to a microservice scenario. Rainbow models the system in terms of \emph{components} (e.g., Clients, Servers, ServerGroups), their \emph{properties} (such as \texttt{Client.responseTime}, \texttt{Component.usedMemory}, or \texttt{ServerGroup.replicas}), 
and \emph{adaptation operators} (such as \texttt{Component.changeParam()} or \texttt{ServerGroup.addServer()})~\cite{1301377}. In a K8s cluster, there is a straightforward mapping between this architectural model and the REST API of K8s: \emph{Server} can be mapped to a K8s Pod, which is the unit of deployment in K8s, and \emph{ServerGroup} to a K8s Deployment that manages a set of identical Pod replicas.  Invoking \texttt{ServerGroup.changeParam()} then corresponds to a  custom API call that adjusts the Deployment resource~\cite{Kubernetes}.

Rules are then expressed as invariants over this architectural model. For instance, Listing \ref{rainbowrule} implements a horizontal autoscaler that scales up and down the number of servers in server group whenever the response time violates a threshold. When the maximum number of servers has been instantiated, however, it can be considered to switch the entire server group to low-power mode, but only if the server group's resources are fully utilized. 
These rules specify when to trigger adaptations and which operators to invoke. Importantly, Rainbow introduces an explicit \emph{planning} step: multiple strategies may be available for a given invariant violation, and the framework evaluates them before committing to execution. This separation of \emph{strategy selection} (planning) from \emph{execution} enables Rainbow to separate deployment of strategies from their activation.

In summary, Rainbow provides strong support for planning, since strategies are explicitly prepared and evaluated before activation. However, its guarantees for system-wide consistency depend on how invariants are formulated. Finally, its reusability across applications remains limited without a standardized platform such as K8s.

\begin{lstlisting}[style=yaml,caption={Rainbow's proposal for specifying invariants over an architectural model, applied to the low-power mode scenario of the Adaptable TeaStore specification}, label={rainbowrule}]{}
invariant (self.responseTime > maxResponseTime) 
    -> responseTimeStrategy(self);

strategy responseTimeStrategy(Client c) {
   ServerGroup sg = getServerGroup(c);
   if sg.replicas < maxReplicas
      then sg.addServer(); 
   else if sg.usedMemory > threshold
      then sg.changeParam("power_mode", "low");
}
\end{lstlisting}

\subsection{Operator pattern}

As stated above, K8s exposes its functionality through a REST API that serves as the central entry point for interacting with the cluster. Every operation -- whether creating, updating, or deleting workloads -- proceeds through this API. Clients submit declarative specifications of the desired state, and K8s ensures that the cluster converges toward this state by invoking dedicated controllers. This design embodies a declarative configuration approach: instead of issuing imperative commands, users declare what they want, and the system continuously works to make it so~\cite{Kubernetes}. 

At the heart of this architecture lies the principle of \emph{level-based reconciliation}. Controllers do not merely react to single events; rather, they periodically inspect the actual cluster state and compute corrective actions necessary to align it with the desired specification. Because the declarative state remains authoritative, controllers are idempotent and self-healing: missed events or transient failures cannot derail adaptation. This contrasts with \emph{event-based reconciliation}, where immediate reactions to notifications risk brittleness and inconsistency when events are delayed or lost.

\begin{figure}[ht]
    \centering
    \includegraphics[width=0.65\textwidth]{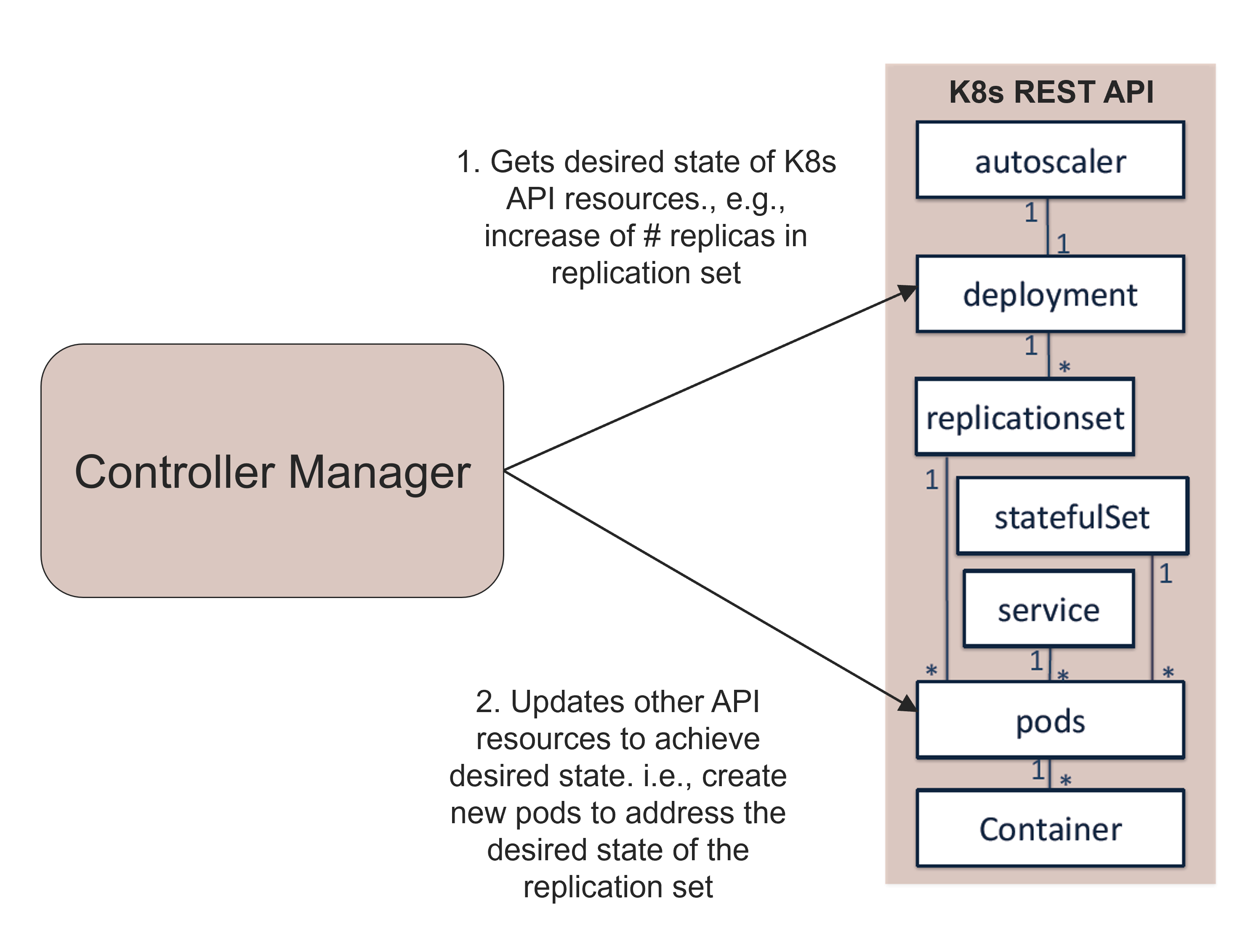}
    \caption{K8s control-plane architecture. The Controller Manager runs as a single process hosting multiple built-in controllers that reconcile the desired and actual state of the K8s API resources.}
    \label{fig:k8s-architecture}
\end{figure}

The built-in controllers for core K8s resources (e.g., Deployments, ReplicaSets, Pods, Services) are packaged together into the \emph{Controller Manager} (cfr. Figure \ref{fig:k8s-architecture}). This component usually runs as a single Pod in the control plane but internally hosts many separate controllers, each dedicated to one API resource. The Controller Manager thus centralizes responsibility for maintaining consistency across all built-in objects, leveraging multi-threading and batching to scale efficiently.

The \emph{Operator pattern}~\cite{k8sOperators} generalizes the Controller Manager architecture to domain-specific applications. Conceptually, it can be described as a threefold approach~\cite{GrasshopperTNSM, buggyoperators, operatorsfailures, operatorssecurity}:
\begin{enumerate}
    \item \textbf{In-cluster API access}: Operators interact directly with the K8s REST API, gaining the same control surface as built-in components.
    \item \textbf{Custom Resource Definitions (CRDs)}: The API itself is extended with new resource types, enabling domain-specific abstractions beyond the default K8s objects.
    \item \textbf{Reconciliation loop}: For each CRD, the operator implements a controller that performs level-based reconciliation, ensuring that the observed state matches the desired state expressed in the custom resource.
\end{enumerate}

\begin{figure}[h]
    \centering
    \includegraphics[width=0.65\textwidth]{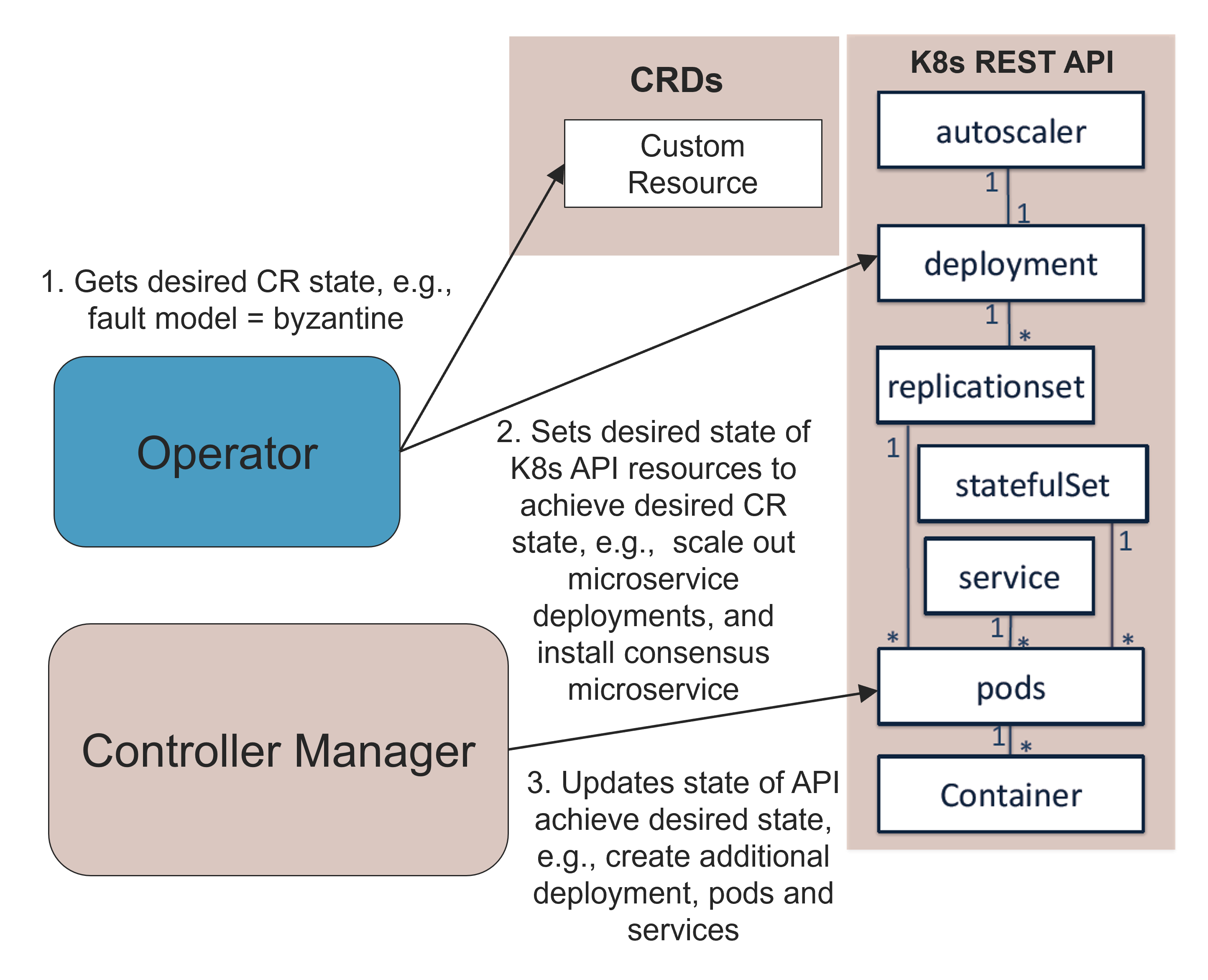}
    \caption{An operator is a dedicated controller for one specific Custom Resource kind. To drive a system to the desired state as specified by a custom resource, it performs updates to the core K8s API resources. This in turn triggers the Controller Manager to inspect what needs to be changed in the system and eventually to drive the system towards the desired configuration.}
    \label{fig:operator-architecture}
\end{figure}

Unlike the Controller Manager, operators are deployed as independent Pods within the cluster (cfr. Figure \ref{fig:operator-architecture}). Each operator encapsulates the logic for a specific domain, such as managing a database or message queue, and scales independently of the core control plane. This decentralization is a defining feature: while the Controller Manager is a single Pod running many built-in controllers, operators distribute control logic into separate workloads, each responsible for reconciling its own CRDs. This design allows operators to encapsulate operational knowledge for any type of domain-specific management concept into a reusable component.

The main strength of the Operator pattern is its scalability, as controllers are isolated and resource type-specific, and reconciliation can proceed in parallel for many resources at once. For example, in the case of Pods, events of each Pod are managed by a separate controller thread. As such there is strong support for system-wide consistency.

However, the pattern also has drawbacks. Developing sophisticated operators requires expertise in controller design. Although operators are meant to be stateless, certain use cases require keeping local state, which introduces concurrency challenges and locking complexity. Existing domain-specific operator libraries are also often buggy or insecure, exposing clusters to potential attacks~\cite{GrasshopperTNSM, buggyoperators, operatorsfailures, operatorssecurity}. Therefore, for simple adaptations, the cost of developing and maintaining a custom operator may outweigh the benefits. Finally, since operators are primarily infrastructure-focused, embedding intricate business logic risks producing bloated and hard-to-maintain systems. In summary, there is weaker support for modular reasoning and evolution, especially when embedding business logic inside an operator.

\subsection{Aspect-Oriented Programming}

Aspect-Oriented Programming (AOP) provides a mechanism to modularize cross-cutting concerns by separating them from the core business logic~\cite{Kiczales1997AOP}. In Aspect/J~\cite{aspectj}, the first and still most popular AOP extension for Java~\cite{baeldungAspectJ}, \emph{pointcuts} define join points in the application where specific actions (\emph{advices}) are woven at run time or compile time around these join points.

Advantages include:
\begin{inparaenum}[(1)] 
    \item separation of concerns: adaptation mechanisms such as monitoring, logging, or fault tolerance can be isolated from core functionalities~\cite{Kiczales1997AOP},
    \item precise event filtering: AOP gives fine-grained control over where and when adaptations are triggered through declarative pointcut expressions~\cite{aspectj},
    \item non-invasive changes: crosscutting behavior variations can be woven into classes without modifying the base source code of these classes~\cite{aspectj}.
\end{inparaenum}

Limitations are:
\begin{inparaenum}[(1)] 
    \item fragile pointcut problem: minor changes in the application's structure (e.g., method renaming) can silently break pointcut bindings, leading to adaptation failures~\cite{fragilepointcuts},
    \item debugging complexity: identifying the origin of dynamic behavior can be difficult, especially in large software systems~\cite{aopdebugging}.
\end{inparaenum}

\subsection{Context-Oriented Programming}

Context-Oriented Programming (COP)~\cite{Hirschfeld2008Context} promotes the idea that the behavior of software entities can vary dynamically based on the execution context. Context-dependent behaviors are organized into \emph{layers}, which can be dynamically activated or deactivated. 

Advantages include:
\begin{inparaenum}[(1)] 
    \item dynamic behavioral variation: COP allows seamless switching between different adaptation strategies depending on the context (e.g., high-load vs low-load behavior)~\cite{Hirschfeld2008Context},
    \item  encapsulation of contextual variations: behavioral variations are cleanly modularized so that code readability and maintainability is improved~\cite{Hirschfeld2008Context},
    \item  fine-grained adaptation: specific functionalities or subsystems can be targeted without impacting unrelated parts of the system~\cite{Hirschfeld2008Context}.
\end{inparaenum}

A key limitation concerns scope management complexity, however, as defining the scope and lifetime of an adaptation activation remains non-trivial, particularly in event-driven or multi-threaded systems~\cite{COPscopes,COPscopes2}. This may cause problems in asynchronous applications or event-based systems. Consider a listener object that registers itself with an event source object while some layers are activated. These layers should be explicitly activated again when the event source invokes the listener object in a separate thread. This requires keeping state on which layers were activated for which listener object, which is cumbersome and complex.

\section{Robust adaptation using the Operator pattern}
\label{sect:operatorpattern}

In previous work, we applied the Operator pattern to implement vendor-agnostic configuration of K8s clusters in a cloud federation \cite{fi15020063}. A desired federation state is specified using a CRD, and a reconciliation loop ensures that the actual state converges to this specification. 

A similar system-wide approach can also be applied to microservice-agnostic adaptation. We outline two possible scenarios that demonstrate how adaptation policies can be implemented in a robust and declarative manner. The CRD extensions of the K8s API provide a persistent, inspectable representation of adaptation state, while the operator enforces convergence toward the declared desired configuration. 

\subsection{Low-power mode adaptation}
Consider an operator that implements a variant of the low-power mode scenario that has been presented in Listing \ref{rainbowrule}. It enables a low-power mode after detecting that the TeaStore WebUI microservice is running out of memory and cannot be further scaled out by adding replicas. To prevent oscillation between power modes, however, the switch should only occur if memory contention is persistent. Moreover, all replicas of the WebUI microservice must be consistently reconfigured to ensure a uniform user exper
\begin{flushleft}

\end{flushleft}
ience. The design of the operator assumes that the WebUI service exposes a REST API for toggling low-power mode and that a Horizontal Pod Autoscaler (HPA) manages the number of replicas within configured bounds according to a specified resource utilization ratio~\cite{Kubernetes}. The maximum replica count thereby serves as a threshold condition for triggering adaptation.   

\begin{figure}[htbp]
\centering
\includegraphics[width=1.0\linewidth]{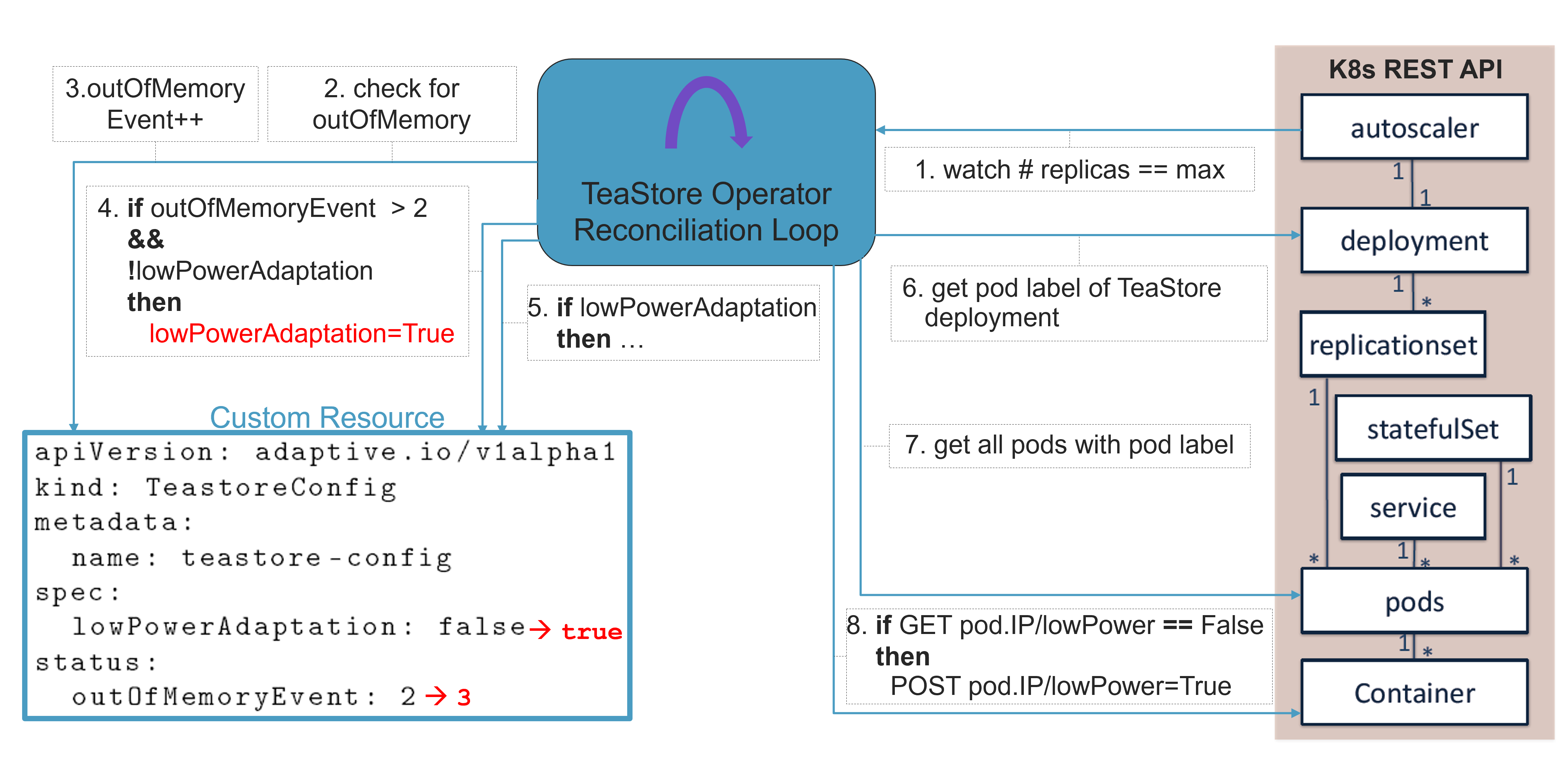}
\caption{The TeaStore operator's reconciliation logic with event aggregation.}
\label{fig:teastoreoperator}
\end{figure}

This functionality can be realized using a CRD and a reconciliation loop. The CRD consists of a boolean field for enabling low-power mode and an integer field for counting out-of-memory (OOM) events (see Listing~\ref{adaptive-config}). The operator executes a single reconciliation loop, as illustrated in Figure~\ref{fig:teastoreoperator}. 

\begin{lstlisting}[style=yaml,caption={TeaStoreConfig Custom Resource Example}, label=adaptive-config]
apiVersion: adaptive.io/v1
kind: TeaStoreConfig
metadata:
  name: teastore-config
spec:
  lowPowerAdaptation: false
  timeInterval: 300 #seconds
status:
  outOfMemoryEvent: 2
  epochStartTimeInterval: 1763744203
\end{lstlisting}

As depicted in Figure~\ref{fig:teastoreoperator}, the reconciliation loop first monitors the HPA configuration until the maximum number of replicas is reached (step~1) and aggregates OOM events over a time interval (steps~2–3). If at least three events occur within the same interval, the operator sets \texttt{lowPowerAdaptation} to \texttt{true} (step~4). Once this field is enabled, the reconciliation logic skips further event aggregation and instead enforces the low-power configuration (steps~5–8). Specifically, it retrieves all Pods belonging to the WebUI Deployment (step~6-7) and invokes their REST API endpoints to enable low-power mode where required (steps~8). Importantly, the reconciliation loop is robust: failed adaptations are detected in subsequent executions, and the operator retries until all replicas are consistently configured.  

This reconciliation-driven design of the operator can be easily implemented in the Kubebuilder framework~\cite{kubebuilder}. Kubebuilder’s controller-runtime periodically invokes the \texttt{Reconcile()} function with an update of all relevant resources (e.g., the TeaStoreConfig CR, the HPA). Multiple responsibilities -- such as aggregating OOM events and enforcing low-power mode -- are then implemented within this function using conditional branching. For example, the function may first count OOM events and decide when to set \texttt{spec.lowPowerAdaptation}; once enabled, the logic bypasses event counting and focuses exclusively on enforcing the desired configuration across Pods. To ensure eventual convergence, the \texttt{Reconcile()} function schedules itself with a certain frequency. Furthermore, Kubebuilder inherently supports parallel reconciliation across different CR instances, enabling scalable adaptation management. An overview of the pseudo-code of the \texttt{Reconcile()} function is provided in Appendix~\ref{app:kube-builder}.

\subsection{Blue-Green deployment of the Recommender service}
Another example where the Operator pattern may be useful is when switching the ML model of the Recommender microservice. If the adaptation logic would immediately set a new Recommender model, the state of the old Recommender model is also immediately discarded. This is not desirable because of two reasons. Firstly, the new model should first be trained before it can be put into commission, thus there should be a dark launch of the new model while the old model keeps serving requests. Secondly, when the new model is trained, it is desirable to defer decommissioning the old version until it has finished processing ongoing requests.  In an operator, setting the Recommender model is a modification of the desired state of a CR. An operator can then implement a blue-green deployment of the Recommender service, where the new service is trained with live traffic without exposing the new service to clients~\cite{bluegreen}. Once the new model is trained, client traffic is switched to the new service and the old service is deleted.

\section{Modularity through AOP and COP}
\label{sect:modularity}

In this section we review some of our previous work on applying AOP and COP to decouple not only the adaptation logic but also to support client-specific activation of behavioral variations. 

\subsection{Weaving the fabric of the control loop using AOP}
\label{sect:eval:aop}

This section shows how AOP can be used to make the original Rainbow framework both generally applicable and reusable across applications. In our previous work~\cite{haesevoets2010weaving}, we presented an approach to constructing self-adaptive systems by combining the ideas behind Rainbow and AOP. To improve the reusability property of Rainbow, we use AOP to modularize the MAPE-K control loop, probes and effectors, and architectural model. The MAPE-K loop is also developed using AOP, with separate aspects for monitoring, analysis, planning, and execution of effectors. 

\begin{figure}[h]
    \centering
    \includegraphics[width=0.8\linewidth]{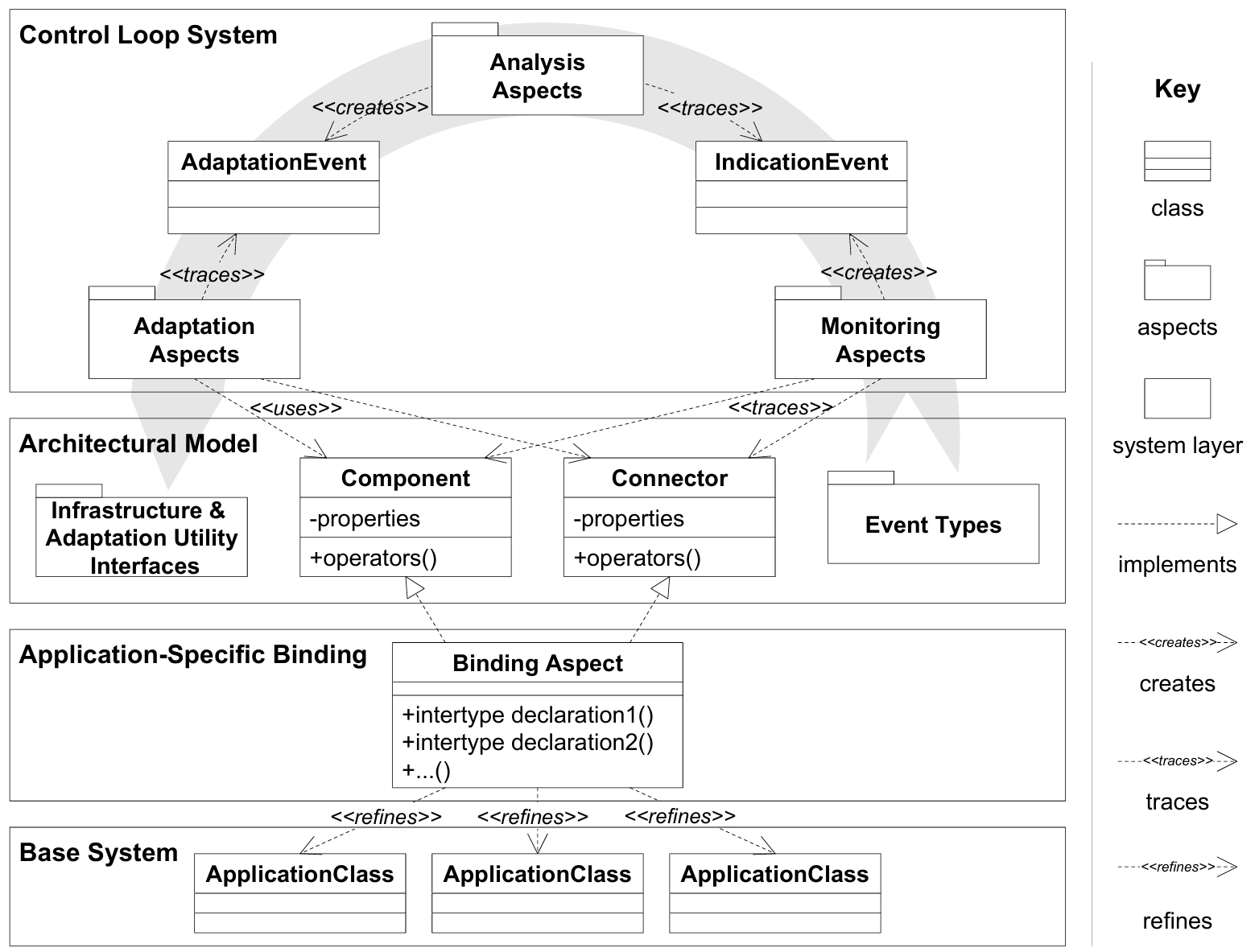}
    \caption{Aspect-oriented realization of the MAPE-K control loop (adapted from~\cite{haesevoets2010weaving}).}
    \label{fig:aop-control-loop}
\end{figure}

Figure~\ref{fig:aop-control-loop} illustrates how each part of the MAPE-K loop is realized through aspects that are woven into different parts of the application, forming an adaptable fabric over the system. The architectural model is bound to the application by means of the inter-type declaration construct of Aspect/J, which allows to extend classes with new interfaces. Pointcuts will convert application-specific events to generic Event objects to which the Monitoring aspects listen. 
For example, the following aspect introduces the architectural abstractions \texttt{Component} and \texttt{Connector} into the concrete classes \texttt{WebClient} and \texttt{Socket}, respectively:

\begin{lstlisting}[style=Java, caption={Binding application classes to model abstractions using inter-type declarations, pointcuts and advice. The \texttt{cflowbelow} construct is intentionally used to select join points that occur within the control flow of the start pointcut, but not the start join point itself.}]
public aspect ClientServerBinding {  

  // Bind interfaces of the application model
  declare parents: WebClient implements Component;      
  declare parents: Socket implements Connector; 

  private Connector WebClient.connector;

  // Pointcuts for creating link wrappers
  pointcut start(WebClient c): 
    call(WebClient.main(..)) && target(c);  

  pointcut connectorCreation(Socket s, WebClient c):   
    execution(Socket.new(..)) && this(s) && cflowbelow(start(c));  

  after (Socket s, WebClient c): connectorCreation(s,c) {    
    ((Component)c).connector = new ConnectorWrapper(c,s);  
  }
}
\end{lstlisting}

Here, the \texttt{declare parents} construct ensures that \texttt{WebClient} and \texttt{Socket} are recognized as \texttt{Com\-po\-nent} and \texttt{Connector} instances at the model level, without modifying their original code. The \texttt{point\-cut} definitions then capture application-specific events such as the creation of sockets, and the associated \texttt{after} advice enriches them with model-level constructs like the \texttt{ConnectorWrapper}. In this way, the system’s runtime behavior is projected onto a stable architectural model that abstracts away from the application details, enabling monitoring and adaptation aspects to interact with this model in a uniform and reusable way. Monitoring aspects observe the Event objects of the application model and evaluate system states using the application model interfaces that have been added to application classes by means of inter-type declarations. When relevant events or states are detected, an IndicationEvent is created. Analysis aspects aggregate multiple of these IndicationEvent objects using the tracematch construct of the \texttt{abc} compiler of Aspect/J, which allows filtering over sequences of joint point occurrences. When a match is found with a sequence of events, an AdaptationEvent object is eventually constructed. Adaptation aspects listen to these AdaptationEvent objects and then modify the system's behavior accordingly. The resulting architecture allows the control logic to be distributed across the system in a loosely coupled, yet effective manner, as the use of AOP allows these concerns to be injected without requiring manual modification of the existing code base of the application.

The evaluation of this work shows that in comparison to an object-oriented version of a case study application, adaptation policies can be implemented more concisely, the resulting implementation code is better modularized and there is less coupling among the different implementation modules.
As the pointcuts for event selection rely on a stable application model, this approach suffers less from the fragile pointcut problem. However, the modularization of adaptation policies into different aspects can complicate global reasoning and validation.

\subsection{Client-specific activation of behavioral variations}

In earlier work~\cite{truyen2012cop}, we explored how Context-Oriented Programming (COP) can support runtime customization in multi-tenant Software-as-a-Service applications. In COP, behavioral adaptations for specific users or runtime situations are encapsulated into \emph{layers}, which can be activated or deactivated on demand based on factors such as user role, subscription tier, or environmental conditions.

COP offers several advantages for self-adaptive microservice systems. First, it supports \emph{modular and reversible adaptations}, since adaptation logic can be defined separately from core functionality. Second, layers can be activated \emph{per request}, allowing the system to react smoothly to runtime context changes. Using this mechanism, context-dependent concerns -- such as UI extensions, pricing models, or data access policies -- can be modularized into layers and consistently activated across application tiers.

However, when COP is implemented as a programming language primarily, mainly local systems are targetted. In distributed microservice architectures, context-sensitive customization must also be coordinated across interacting services by middleware components. 

For example, Listing~\ref{rainbow-context-sensitiverule} shows a simplified architectural invariant that enables a low-power mode only for clients with slow network connections. The goal is to compensate for high network round trip latency by allowing the microservice to respond more quickly by running in low-power mode. The control loop monitors both the microservice’s response time and the client’s round-trip latency. When both exceed acceptable thresholds, a \emph{customization property} \texttt{(`power\_mode', `low')} is added to the Connector object associated with the client (see Section~\ref{sect:eval:aop}).

\begin{lstlisting}[style=yaml,caption={Rainbow's proposal for specifying invariants over an architectural model, applied to the low-power mode scenario of the Adaptable TeaStore specification}, label={rainbow-context-sensitiverule}]{}
invariant (self.responseTime > maxResponseTime) 
    -> responseTimeStrategy(self);
strategy responseTimeStrategy(Component c) {
    if c.connector.roundTripLatency > maxRoundTripLatency
        then c.connector.setProperty(`power_mode', `low');
}
\end{lstlisting}

The remaining question is how setting this customization property triggers activation of the corresponding layer in the relevant microservices. Here, our previous work on the Lasagne middleware~\cite{truyen2001} becomes relevant. Each client request that passes through the connector has its header augmented with the customization property by an interceptor. On the service side, another interceptor inspects this property to determine whether the low-power mode must be activated for that specific request.

We originally proposed this mechanism -- propagating customization properties through request headers and interpreting them as triggers for activating modular extensions -- in work predating the COP paradigm~\cite{truyen2001}. It has been implemented as reflective middleware~\cite{truyen2001, truyen2002, Truyen2004DynamicContextSensitiveComposition} and as a Java extension~\cite{JorgensenT03}. Lasagne’s design closely aligns with COP: layers are implemented as sets of wrappers around the base classes in each service, and each layer has a unique identifier. This identifier is associated with the thread handling a request or propagated with message flows via reflection. The active configuration for a request is therefore defined by the set of layer identifiers attached to it and enforced by a layer-aware message dispatcher. If no layer identifiers are present, execution defaults to the base classes.

Lasagne also includes a mechanism to maintain consistent layer activation among collaborating microservices. Customization properties and layer identifiers attached to a request are automatically forwarded as the request flows through the system. Alternatively, microservices may re-interpret these properties independently, preserving their decoupled and self-contained nature. In this configuration, only the client-side customization properties are propagated, and each service uses its own interceptors to interpret them.

Finally, Lasagne addresses a well-known challenge in COP: managing behavioral scope in asynchronous systems. It does this by attaching layer identifiers to object references. When a listener object is passed to an event source object, the listener's reference is annotated with the current set of layer identifiers. When a callback later occurs on that listener, the callback inherits the same identifiers that were active when the listener was registered. This ensures consistent activation of behavioral variations across asynchronous interactions~\cite{Truyen2004DynamicContextSensitiveComposition}.

\section{Discussion}
\label{sect:conclusion}

This paper has explored multiple approaches for decoupling adaptive control from the Adaptable TeaStore microservices. Our central concern has been how well these approaches realize three key self-adaptation properties: \emph{system-wide consistency}, \emph{planning}, and \emph{modularity}.

\textbf{System-wide consistency.} Both the Operator pattern and architecture-based frameworks such as Rainbow provide strong support for enforcing adaptations consistently across container replicas. For example, in the low-power mode scenario, an Operator ensures that all WebUI replicas eventually converge to the same state. By contrast, AOP and COP are primarily concerned with modular integration at the code level and offer no built-in guarantees of consistent adaptation across distributed replicas of the same microservice.

\textbf{Planning.} Planning requires the ability to defer or sequence adaptations until appropriate conditions are met. Rainbow naturally supports this property by distinguishing between planning and execution of adaptation strategies. The Operator pattern can also encode planning policies, as illustrated by the aggregation of OOM event and the dark launch of a new Recommender ML model, but this requires additional engineering effort. AOP and COP support a degree of planning through event filtering and layer activation, but these mechanisms operate at a finer granularity and are harder to coordinate system-wide.

\textbf{Modularity.} Modularity of business logic that is akin to specific microservice themselves is best supported by programming language techniques. AOP enables cross-cutting adaptation logic to be isolated from the application code base, facilitating reuse across services. COP modularizes behavioral variations into layers that can be dynamically activated or deactivated, supporting fine-grained customization. By contrast, the Operator pattern tends to mix domain-specific control logic with infrastructure management code, which can complicate maintenance.

\paragraph{Applicability to the Adaptable TeaStore scenarios.}
The four approaches considered in this paper can be applied in complementary ways to the scenarios defined in the Adaptable TeaStore specification. For scenarios that require \emph{system-wide reconfiguration} -- such as switching all Recommender replicas to low-power mode, falling back to local static services during database or provider outages, or redeploying services after a regional failure -- the Operator pattern is the most suitable option. Its reconciliation loop ensures consistent configuration across replicas and supports planning steps such as dark launches or staged activation. 

Architecture-based adaptation also supports these scenarios by expressing high-level invariants and strategies. In its traditional form, such an
approach requires additional infrastructure (custom probes, effectors, an architectural model, and a dedicated controller). Yet this requirement is
significantly reduced when the architectural layer is implemented on top of K8s  -- as demonstrated by Kubow -- since it can reuse the K8s API,
resource abstractions, and reconciliation behaviour rather than reimplementing these mechanisms from scratch.

For \emph{fine-grained behavioural variation inside a microservice}, such as switching algorithms or enabling a maintenance mode, \emph{AOP} offers clean separation of adaptation logic without modifying the base code. For \emph{context-specific or per-request variations}, for instance activating low-power behaviour only for clients with slow connections, \emph{COP} provides dynamic activation of behavioural layers. However, both AOP and COP operate locally and therefore do not guarantee system-wide consistency.

Overall, a multi-tiered architecture is recommended: Operators handle macro-level adaptation and global consistency, while AOP or COP refine intra-service behaviour at the micro level. Architecture-based approaches are beneficial when explicit planning and strategy selection are required.

\paragraph{Missing elements in the specification.}
Although the reference implementation of Adaptable TeaStore exposes a set of REST endpoints for metrics and offers POST-based adaptation actions, the overall document (specification + reference implementation) still lacks several elements that are necessary to uniformly support the four adaptation approaches discussed in this paper. First, while services provide operational metrics and coarse-grained lifecycle states (e.g., \texttt{RUNNING}, \texttt{TRAINING\-_DATA}, \texttt{GENERATING\-_IMAGES}), there is no specification-level contract that defines how a service must expose its \emph{current adaptation configuration}. In particular, there is no standardised GET endpoint that returns the logical adaptation state of a service (such as whether the Recommender is currently in low power mode, whether caching is enabled, which image provider flavour is active, or whether the circuit breaker is enabled). The POST-based adaptation actions allow these modes to be set, but the document does not prescribe how services should report which of these modes is currently in effect. This is however critical for the reconciliation loop of the Operator pattern which compares actual system state against the desired state. 

Secondly, some of concrete adaptation actions are clearly event-based (e.g., \texttt{Database\-Unavailable\-EventBroadcast}, \texttt{DDoS\-Attack\-Event\-Broadcast}). The specification does not define the \emph{semantics} of these events at the specification level. Missing are precise definitions of: (i) when an event is raised (e.g., what constitutes ``database unavailable'' -- a failed TCP connection, repeated query timeouts, or an application-level exception?), (ii) the expected payload or information content of the event. Without these semantics, external controllers (e.g., operators or architectural controllers) cannot rely on consistent interpretation of triggers across implementations.

Thirdly, no mechanism is standardised for propagating context information -- such as client-specific customization properties -- across microservices. COP techniques require an explicit, specification-level protocol for attaching, forwarding, and interpreting context fields (e.g.\ via HTTP headers), as well as rules for managing context across asynchronous callbacks. This is necessary for ensuring that per-request behavioral variations remain coherent in a distributed setting. 

\section*{Acknowledgments}
This research is partially funded by the Research Fund KU Leuven, and by the Cybersecurity Research Program Flanders. We thank the anonymous reviewers of the WACA 2025 PC for their helpful comments and suggestions.

\bibliographystyle{eptcs}

\bibliography{b}

\appendix

\section{Implementation of level-based reconciliation in Kubebuilder}
\label{app:kube-builder}
The controller logic is written in the form of a reconciliation loop that counts the number of out of memory events and only sets the lowPowerAdaptation to \texttt{true} after one or multiple of these events. Listing \ref{listing-kube-builder1} shows how this can be implemented in Kubebuilder, a popular operator library for K8s \cite{kubebuilder}. A clear disadvantage is the low-level code for maintaining the number of outOfMemory events during a time interval (lines 9-26) and the performance overhead for storing the count and the time interval in the \texttt{Status} object of the \texttt{TeaStoreConfig} CR (line 15). This event aggregation cannot be fully encapsulated in a \texttt{predicate.Predicate}~\cite{kubebuilder_predicates} for event filtering as a \texttt{Predicate} cannot mutate the Status object. 

\mbox{}

\begin{lstlisting}[style=gocode,caption={Controller reconciliation logic with event counting},label={listing-kube-builder1}]
func (r *TeaStoreConfigReconciler) Reconcile(ctx context.Context, req ctrl.Request) (ctrl.Result, error) {
    var config adaptivev1.TeaStoreConfig
    if err := r.Get(ctx, req.NamespacedName, &config); err != nil {
        return ctrl.Result{}, client.IgnoreNotFound(err)
    }
		

    // Simulated external event detection
    if detectOutOfMemory() {
        if currentTimeInEpochs() > config.Status.EpochStartTimeInterval + config.Spec.TimeInterval {
            config.Status.OutOfMemoryCount=0
            config.Status.EpochStartTimeInterval=currentTimeInEpochs()
        }
        config.Status.OutOfMemoryCount++
        if err := r.Status().Update(ctx, &config); err != nil {
            return ctrl.Result{}, err
        }
    }

    // Trigger adaptation if threshold is reached
    if config.Status.OutOfMemoryCount >= 3 && !config.Spec.LowPowerAdaptation {
        config.Spec.LowPowerAdaptation = true
        if err := r.Update(ctx, &config); err != nil {
            return ctrl.Result{}, err
        }
        log.Log.Info("Enabled low power adaptation due to repeated out of memory events.")
    }
		
		[... other code of Listing 6]

\end{lstlisting}

\begin{lstlisting}[style=gocode,caption={Controller reconciliation logic for activating low power mode}, label=listing-kubebuilder2]
		[...previous code of Listing 5]

    // Reconcile actual microservice state with desired low power mode
    if config.Spec.LowPowerAdaptation {
        var podList corev1.PodList
        if err := r.List(ctx, &podList, client.MatchingLabels{"app": "teastore"}); err != nil {
            return ctrl.Result{}, err
        }

        for _, pod := range podList.Items {
            // Query current power state (optional)
            svc := pod.Status.PodIP // or pod.Name if using DNS
            resp, err := http.Get(fmt.Sprintf("http://%s/adapt/lowpowermode", svc))
            if err != nil || resp.StatusCode != http.StatusOK {
						    log.Log.Error(err, "Failed to query low power state", "service", svc)
                continue
            }

            var result struct {
                LowPowerEnabled bool `json:"low_power_enabled"`
            }
            json.NewDecoder(resp.Body).Decode(&result)

            if !result.LowPowerEnabled {
                // Reconcile by enabling low power mode
                http.Post(fmt.Sprintf("http://%s/adapt/lowpowermode/enable", svc), "application/json", nil)
                log.Log.Info("Reconciled low power mode for", "service", svc)
            }
        }
    }

    return ctrl.Result{RequeueAfter: time.Minute}, nil
}
\end{lstlisting}

Setting lowPowerAdaptation to True then triggers the same controller to bring the actual state of the relevant microservices into this desired state. It detects which microservices are still running in high-power mode and changes these to low-power mode (see Listing \ref{listing-kubebuilder2}, lines 10-29 ). Note that the reconciliation loop is robust for failed reconfiguration by rescheduling the controller again (cf. Line 32). Thus, in case some microservices are still running in high-power mode after a first reconfiguration attempt, this will be automatically detected by the controller, which then retries setting them to low-power mode. Note that a modified version of the \texttt{adapt} API endpoint of the Adaptable TeaStore may be needed in order to support a GET of the current state of the adaptation.

\end{document}